\documentclass{aa}

\usepackage[]{graphicx}
\usepackage[]{exscale}

\def\astroph#1#2{#1, {\bf astro-ph}/#2}
\def\aj#1#2#3{#1,      {AJ, }{\bf#2}, #3}
\def\aa#1#2#3{#1,      {A\&A, }{\bf#2}, #3}
\def\aas#1#2#3{#1,     {A\&AS, }{\bf#2}, #3}
\def\apj#1#2#3{#1,     {ApJ, }{\bf#2}, #3}
\def\mn#1#2#3{#1,      {MNRAS, }{\bf#2}, #3}
\def\Nature#1#2#3{#1,  {Nature, }{\bf#2}, #3}
\def\book#1#2#3{#1, #2 (#3)}
\def\proceeding#1#2#3#4#5{#1, in #2, #3 (#4), #5}

\begin{document}

\title{On the Ionisation Fraction in Protoplanetary Disks II: \\
The Effect of Turbulent Mixing on Gas--phase Chemistry}

\author{Martin~Ilgner and Richard~P.~Nelson}


\institute{Astronomy Unit, Queen Mary, University of London, 
Mile End Road, London E1 4NS, U.K.}

\date{Received 21 July 2005 / Accepted 30 August 2005}

\titlerunning{Ionisation fraction in disks}

\authorrunning{M.~Ilgner, R.P.~Nelson}

\abstract{We calculate the ionisation fraction in protostellar disk 
models using two different gas--phase chemical networks, and examine 
the effect of turbulent mixing by modelling the diffusion of chemical 
species vertically through the disk. The aim is to determine in which 
regions of the disk gas can couple to a magnetic field and sustain 
MHD turbulence. The disk models are conventional $\alpha$--disks, and 
the primary source of ionisation is X--ray irradiation from the central 
star. We assume that the vertical mixing arises because of turbulent 
diffusion, and accordingly equate the vertical diffusion coefficient, 
${\cal D}$, with the kinematic viscosity, $\nu$.\\
We find that the effect of diffusion depends crucially on the elemental 
abundance of heavy metals (magnesium) included in the chemical model. 
In the absence of heavy metals, diffusion has essentially no effect on 
the ionisation structure of the disks, as the recombination time scale 
is much shorter than the turbulent diffusion time scale. When metals 
are included with an elemental abundance above a threshold value, the 
diffusion can dramatically reduce the size of the magnetically 
decoupled region (``dead zone''), or even remove it altogther. This 
arises when recombination is dominated by metal ions, and the 
recombination time exceeds the vertical diffusion time scale. For a 
complex chemistry the elemental abundance of magnesium required to 
remove the dead zone is $x_{\rm Mg} = 10_{}^{-10}$--$10_{}^{-8}$. We 
also find that diffusion can modify the reaction pathways, giving rise to 
dominant species when diffusion is switched on that are minor species when 
diffusion is absent. This suggests that there may be chemical signatures 
of diffusive mixing that could be used to indirectly detect turbulent 
activity in protoplanetary disks.\\
We find examples of models in which the dead zone in the outer disk region 
is rendered deeper when diffusion is switched on. This is caused by turbulent 
mixing diluting the electron fraction in regions where the ionisation degree 
is marginally above that required for good coupling.\\
Overall these results suggest that global MHD turbulence in protoplanetary 
disks may be self--sustaining under favourable circumstances, as turbulent 
mixing can help maintain the ionisation fraction above that necessary to 
ensure good coupling between the gas and magnetic field.

}

\maketitle

\section{Introduction}
Observational studies of young stars in star forming regions indicate 
that protostellar disks are a common occurrence (e.g. Beckwith \& 
Sargent 1996; Prosser et al. 1994). Many disks show signatures of active 
accretion, with the probability of accreting disks being present, and 
the appparent gas accretion rate, scaling inversely with the age of the 
stellar system in which the young stars are embedded. The canonical value 
for the gas accretion rate, however, is often quoted as being
${\dot M} \sim 10_{}^{-8}$ M$_{\odot}$ yr$_{}^{-1}$ (e.g. Sicilia--Aguilar 
et al. 2004).\\
\indent
There are a number of potential mechansims that may lead to angular 
momentum transport in protostellar disks, giving rise to the observed mass 
accretion. Angular  momentum transport that occurs globally throughout the 
disk, producing accretion at the observed rates, probably requires turbulence 
to act as a source of anomalous viscosity. At the present time the only 
source of turbulence in accretion disks that is known to work is the 
magnetorotational instability (MRI), which gives rise to MHD turbulence 
(Balbus \& Hawley 1991; Hawley \& Balbus 1991; Hawley, Gammie \& Balbus 
1996).\\
\indent
As has been well documented in the literature, there are continuing 
questions about the applicability of the MRI to protostellar disks because 
of their high densities and low temperatures that lead to low levels of 
ionisation (e.g. Blaes \& Balbus 1994; Gammie 1996). Magnetohydrodynamic 
simulations of disks including ohmic resistivity (Fleming, Stone \& Hawley 
2000) show that for magnetic Reynolds numbers ${Re}_{\rm m}^{}$ smaller than 
a critical value ${Re}_{\rm m}^{\rm crit}$, turbulence cannot be sustained 
and the disks return to a near--laminar state. Typically the gas--phase 
electron fraction should be $x[\mathrm{e}_{}^-] \simeq 10_{}^{-12}$ for 
disks to be able to sustain MHD turbulence.\\
\indent
A number of studies of the ionisation fraction in protostellar disks have 
appeared in the literature. Gammie (1996) suggested that disks may have 
magnetically ``active zones" sustained by thermal or cosmic ray ionisation, 
adjoining regions that were magnetically inactive -- ``dead zones". More 
recent studies have examined this issue in greater depth. Sano et al. (2000) 
used a more complex chemical model that included dust grains. Glassgold et 
al. (1997) and Igea et al. (1999) examined the role of X--rays as a source of 
ionisation in protoplanetary disks. Fromang et al. (2002) considered the role 
of heavy metals in determining the ionisation fraction because of the 
potential importance of charge--transfer reactions. Matsumura \& Pudritz 
(2003) examined the ionisation fraction in the externally heated, passive 
disk model proposed by Chiang \& Goldreich (1997) using the Sano et al. (2000) 
chemical reaction network. Semenov et al. (2004) recently studied disk 
chemistry using a complex reaction set drawn from the UMIST database.\\
\indent
In a recent paper (Ilgner \& Nelson 2005 -- hereafter ``paper 1''), we 
examined and compared the predictions made by a number of chemical reaction 
networks about the ionisation fraction in standard $\alpha$--disk models. 
This study included an examination of the reaction scheme proposed by 
Oppenheimer \& Dalgarno (1974), and more complex schemes drawn from the UMIST 
database, in addition to a number of gas--grain chemical networks. In this 
paper we extend this initial study to examine the role of turbulent mixing in 
determining the ionisation fraction in protoplanetary disk models using gas--phase 
chemistry. We consider the simple Oppenheimer \& Dalgarno (1974) model, and a 
more sophisticated chemical reaction network based on the UMIST database. We 
allow the vertical diffusion of each chemical species to occur, equating the 
diffusion coefficient ${\cal D}$ to the kinematic viscosity $\nu$ that drives 
disk accretion. We find that the inclusion of diffusion can have a very 
significant effect on the ionisation structure of protoplanetary disks, in 
particular when a small abundance of heavy metals (magnesium) is introduced 
into the reaction networks. In some cases the disk can modify the ionisation 
fraction sufficiently that the dead zone disappears entirely.\\
\indent
This paper is organised as follows. In section~\ref{section2} we describe the 
chemical models, our implementation of the diffusion equation for chemical 
species, and the numerical method used to solve the reaction--diffusion equations. 
In section~\ref{results} we discuss the various models that we compute and present 
their results. In section~\ref{summary} we summarise our main findings and discuss 
their potential consequences for protoplanetary disks.

\section{Reaction--diffusion model}
\label{section2}
We consider a system of $s$ chemical species whose local abundances in 
the protoplanetary disk evolve due to chemical reactions and vertical 
diffusive transport, arising because of concentration gradients and 
driven by turbulence. In the discussion below we define $r$ to be the 
number of chemical reactions, and $n$ to be the total molar density 
(expressed in units of mols cm$^{-3}$). The global elemental composition 
of the system is conserved by applying appropriate boundary conditions 
in our reaction--diffusion model.

\subsection{Disk models}
\begin{figure}[t]
\includegraphics[width=.50\textwidth]{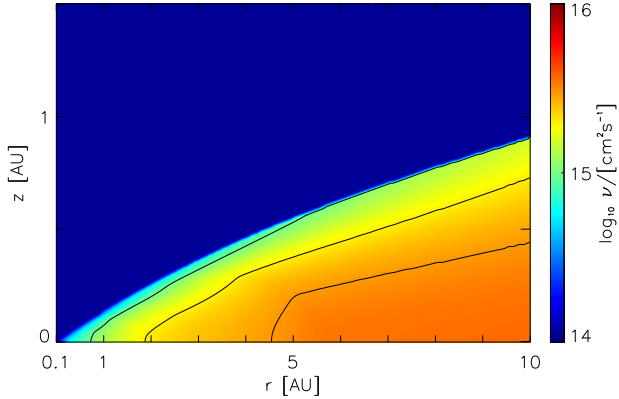}
\caption{The kinematic viscosity $\nu$; the contour lines refer to 
values $1 \times 10^{15}, 2 \times 10^{15}, 3 \times 10^{15} \ 
\mathrm{cm^2s^{-1}}$. The disk parameters are $\alpha = 10^{-2}$ and 
$\dot{M} = 10^{-7} \ \mathrm{M_{\odot} yr^{-1}}$.}
\label{kinematic_viscosity1}
\end{figure}

\begin{figure}[t]%
\includegraphics[width=.50\textwidth]{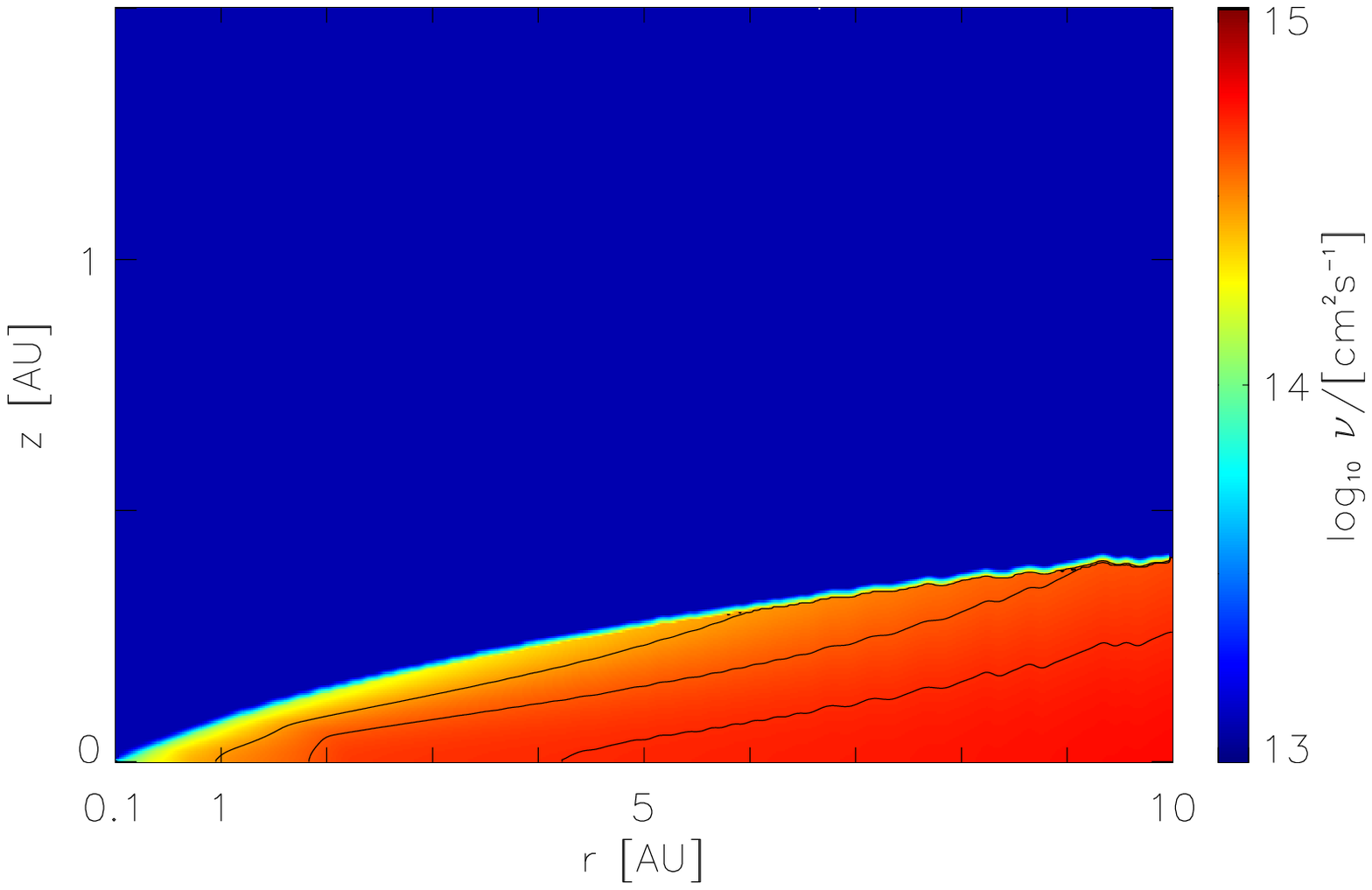}
\caption{The kinematic viscosity $\nu$; the contour lines refer to 
values $3 \times 10^{14}, 4 \times 10^{14}, 5 \times 10^{14} \ 
\mathrm{cm^2s^{-1}}$. The disk parameters are $\alpha = 5 \times 10^{-3}$ 
and $\dot{M} = 10^{-8} \ \mathrm{M_{\odot} yr^{-1}}$.}
\label{kinematic_viscosity2}
\end{figure}

The underlying disk models considered are standard $\alpha$--disks.
Details are given in paper 1 and references therein. To recap: the 
disks are assumed to orbit a young solar mass star and undergo viscous 
evolution. We use the $\alpha$ prescription for the viscous stress, such 
that the kinematic viscosity $\nu= \alpha c_s^2/\Omega$, where $c_s$ is 
the sound speed and $\Omega$ is the local Keplerian angular velocity. 
Heating of the disk is provided by viscous dissipation alone, and cooling 
by radiation transport in the vertical direction. The disk structure is 
obtained by solving for hydrostatic and thermal equilibrium. We employ 30 
zones in the vertical direction (from midplane to disk surface) and 100 
zones in radius between $0.1 \le R \le 10$ AU when computing the chemical 
evolution. The underlying disk model was computed using 100 cells in the 
vertical direction, and values were interpolated onto the 30 grid points 
used in computing the chemistry. The disk models are completely specified 
by the mass accretion rate, ${\dot M}$ and the value of $\alpha$. We 
consider two models: one with ${\dot M}=10_{}^{-7}$ M$_{\odot}$ yr$_{}^{-1}$
and $\alpha=10_{}^{-2}$, the other with ${\dot M}=10_{}^{-8}$ M$_{\odot}$ 
yr$_{}^{-1}$ and $\alpha=5 \times 10_{}^{-3}$. The mass contained in these 
models is 0.0087 M$_{\odot}$ and 0.0049 M$_{\odot}$ respectively. Contour 
plots showing the distribution of the kinematic viscosity, $\nu$, are shown 
in figures~\ref{kinematic_viscosity1} and \ref{kinematic_viscosity2}.

\subsection{Kinetic model}
\label{sec:kinetic}
We have applied two different kinetic models for the gas--phase chemistry. 
The first is a simple five component model introduced originally by 
Oppenheimer \& Dalgarno (1974). The second is a much more sophisticated 
reaction scheme based on the UMIST database (Le Teuff et al. 2000). We 
assume that ionisation of the disk material arises because of incident 
X--rays that originate in the corona of the central T Tauri star. We 
neglect contributions from Galactic cosmic rays as they are not expected 
to penetrate into inner disk regions we consider due to the stellar wind. 
The details of these models have been described in paper 1, where they 
were given the labels \texttt{model1} and \texttt{model3}. We maintain 
this labelling scheme in this paper so as to establish continuity with 
our previous work. \\
\indent
The kinetic equations obey the law of atomic balance:
\begin{equation}
\sum_{i=1}^{s} \nu_{ij}^{} = 0, \ (j=1,2,\dots, r).
\label{eq:1}
\end{equation}
Here $\nu_{ij}$ divided by the molecular mass $M_i^{}$ of the 
$i^{\mathrm{th}}$ component (defined in units of grams/mol) is 
proportional to the stoichiometric coefficient, while the 
$i^{\mathrm{th}}$ species is involved in the $j^{\mathrm{th}}$ 
chemical reaction. The molar density, $n_i^{}$, of the 
$i_{}^{\mathrm{th}}$ component is related to the mass density, 
$\rho_i^{}$, of the $i_{}^{\mathrm{th}}$ component through
\begin{equation}
\rho_i^{} = M_i^{}n_i^{}.  
\label{eq:2}
\end{equation}
The rate of change of the molar density of the $i_{}^{\mathrm{th}}$ 
component within a given volume due to chemical reactions and flow of 
species $i$ into that volume is 
\begin{equation}
\frac{\partial n_i^{}}{\partial t} = 
- \nabla \cdot \left( n_i^{} {\bf v}_i^{} \right) 
+ \sum_{j=1}^r \nu_{ij}^{} J_j^{}.
\label{eq:3}
\end{equation}
$J_j^{}$ denotes the chemical reaction rate associated with the 
$j_{}^{\mathrm{th}}$ chemical reaction, while $\nu_{ij}^{} J_j^{}$ 
is the formation/destruction rate of the $i_{}^{\mathrm{th}}$ 
component due to the $j_{}^{\mathrm{th}}$ chemical reaction. 
${\bf v_i^{}}$ is the velocity of the $i_{}^{\mathrm{th}}$ component. 

\subsection{Diffusion}
\label{sec:diffusion}
The first term on the right hand side of equation~(\ref{eq:3}) represents 
the rate of flow of species $i$ into a given volume. In this paper we 
assume that the underlying protoplanetary disk is turbulent, and that 
turbulent diffusion acts to transport chemical species from one region of 
the disk to another. As we are primarily interested in how diffusion affects 
the ionisation fraction of the disk material, and we expect the largest 
gradients in the electron fraction to be in the vertical ($z$) direction, 
we consider only vertical diffusion in the paper. Equation~(\ref{eq:3}) 
becomes
\begin{equation}
\frac{\partial n_i^{} }{\partial t}  = 
\frac{\ \partial}{\partial z} \left( n {\cal D} 
\frac{\ \partial}{\partial z} x_i^{} \right) + 
\sum_{j=1}^r \nu_{ij}^{} J_j^{}, \ (i=1,\dots, s)
\label{eq:14}
\end{equation}
where ${\cal D}$ is the turbulent diffusion coefficient, and $x_i=n_i/n$ 
is the fractional abundance of species $i$. We present a fuller discussion 
of the derivation of equation~(\ref{eq:14}) in appendix. A similar set of 
equations has been used by numerous authors who have examined the effect of 
turbulent diffusion on the chemical state of molecular clouds (Xie, Allen 
\& Langer 1995; Rawlings \& Hartquist 1997; Willacy, Langer \& Allen 2002) 
and on the role of radial mixing in protoplanetary disks (Wehrstedt \& Gail 
2002, 2003). In this work we adopt the approximation that ${\cal D}= \nu$, 
where $\nu$ is the kinematic viscosity that drives the radial diffusion of 
mass through the protostellar accretion disk. Although there is still some 
debate about the precise relationship between the vertical diffusion 
coefficient and the effective kinematic viscosity that arises because of MHD 
turbulence in astrophysical disks (see Carbillado, Stone \& Pringle 2005; 
Johansen \& Klahr 2005), it is not expected that ${\cal D}/\nu$ differs 
greatly from unity. 
As already discussed, the basic kinetic models that we consider in this 
paper are models \texttt{model1} and \texttt{model3} described in paper 1. 
When these kinetic models are used in conjunction with turbulent mixing 
we use the labels \texttt{model1D} and \texttt{model3D}, respectively. 
When diffusion is switched off we often use the term ``pure kinetic models'' 
when describing the results in later sections. 

\subsection{Numerical method}
\label{sec:numerical}
The reaction--diffusion model is governed by the set of $s$ coupled 
parabolic partial differential equations~(\ref{eq:14}). These equations 
can be interpreted as a linear superposition of two operators, one 
describing the mass transfer of species $i$ due to diffusion, and the 
other describing changes due to chemical reactions. There are number of 
possible approaches to solving these equations. We employed the method 
of lines, which is a technique used, e.g., in atmospheric physics that 
solves the full system of equations simultaneously (e.g. Chang et al. 
1974), rather than a more usual operator splitting approach. Hence, we 
transformed the system of $s$ PDEs into a system of $s \times nz$ ODEs, 
where $nz$ is the number of grid points in the $z$ direction. Adopting 
a uniform mesh along this coordinate direction, we replaced the spatial 
derivative in equation~(\ref{eq:14}) by a finite differencing scheme.\\
\indent
The boundary conditions are taken to be symmetric at $z=0$ (disk midplane) 
and no flux at $z=H$ (photospheric disk height) for all times $t$. Applying 
the Crank--Nicholsen finite differencing scheme and the chosen boundary 
conditions satisfy the conservation of elements and charges at any 
cylindrical radius $R$. Using $nz = 30$, we found that the elements and 
charges are conserved at all cylindrical radii and for all times $t$. At 
$t \le 10_{}^{5} \rm \ yrs$, for example, the change in both the elemental 
abundances and the total charge was below $10_{}^{-8}$ per cent compared 
with the corresponding values at $t=0 \rm  \ yr$.\\
\indent
We have allowed all species, including molecular hydrogen and helium, to 
be mixed by diffusion. We thus checked {\em a posteriori} if the assumed 
constancy of the mean molecular weight $\mu = 2.33$ is satisfied. We 
find that it is, as expected from the fact that the value of $\mu$ is 
dominated by hydrogen and helium, and these species do not develop 
significant concentration gradients, thus precluding significant diffusion.\\
\indent
We also allowed free electrons to be mixed by diffusion. Differences in the 
concentration gradients of the ions and electrons can in principle lead to 
different diffusion velocities which may result in a non--zero diffusion flux 
of charges per zone. We found, however, that this did not occur.

\subsection{Initial conditions}
The reaction--diffusion calculations are initiated with all chemical elements 
being in neutral atomic form, apart from hydrogen which is assumed to be in 
molecular form. The species were distributed in the disk uniformly. Details are 
given in paper 1. Chemical changes lead to concentration gradients that then 
initiate the action of diffusion.

\section{Results}
\label{results}
We have evolved the disk chemistry using the kinetic models described in 
section~\ref{sec:kinetic}, including mixing due to turbulent diffusion. The 
primary aim of this work is to compare and understand the differences in 
the distribution of free electrons that arise in the different models, and 
in particular to understand the role that turbulent mixing plays in 
determining the free electron fraction and distribution. We wish to determine 
which parts of the disk are sufficiently ionised for the gas to be well 
coupled to the magnetic field, and thus able to maintain MHD turbulence, and 
which regions are too neutral for such turbulence to be maintained. We refer 
to those regions as being ``active'' and ``dead'' zones respectively, with 
the region bordering the two being the ``transition'' zone. The important 
discriminant that determines whether the disk is active or dead is the 
magnetic Reynolds number, ${Re}_{\rm m}^{}$, defined by
\begin{equation}
{Re}_{\rm m}^{} = \frac{H c_s}{\mu_{\rm m}^{}}
\label{reynolds}
\end{equation}
where $H$ is the disk semi--thickness, $c_s$ is the sound speed, and $\mu_{\rm m}^{}$ 
is the magnetic diffusivity (not to be confused with the mean molecular 
weight). Numerical simulations (e.g. Fleming, Stone \& Hawley 2000) indicate 
that a critical value of the magnetic Reynolds number, ${Re}_{\rm m}^{\rm crit}$, 
exists such that non linear MHD turbulence cannot be sustained if ${Re}_{\rm m}^{}$ 
falls below ${Re}_{\rm m}^{\rm crit}$. We adopt a value of 
${Re}_{\rm m}^{\rm crit} = 100$ in this paper, following the value used in 
paper 1 and Fromang et al. (2002). We are able to calculate the distribution of 
${Re}_{\rm m}^{}$ within our disks. Regions with ${Re}_{\rm m}^{} < 100$ are deemed to 
be magnetically dead, and those with ${Re}_{\rm m}^{} > 100$ magnetically active.\\
\indent
When it comes to diffusion we consider the two extremes with the diffusion 
coefficient ${\cal D} = \nu$ and  ${\cal D} = 0$. Models with ${\cal D}=\nu$ 
are labelled \texttt{model1D} and \texttt{model3D}. Models with ${\cal D}=0$ 
reduce to the pure kinetic models \texttt{model1} and \texttt{model3} already 
discussed in paper 1.\\
\indent
We solved the equations by integrating over a time interval of 100,000 yrs. 
Hence, the ionisation fraction $x[\mathrm{e}^-]$ is a function of time $t$, 
and in principle so is the location of the transition zone. For all kinetic 
models, however, the change in the vertical location of the transition zone 
at all cylindrical radii in the computational domain was below the grid 
resolution for $t > 50,000 \ \mathrm{yrs}$.

\subsection{Diffusion versus Recombination}
\label{diffusion+recombination}
In order for diffusion to modify the ionisation fraction of material near 
the transition zone in protoplanetary disks, and hence modify the size of 
the dead zone, the vertical diffusion time should be less than the 
recombination time for free electrons.\\
\indent
We consider diffusion across a scale height at $R \simeq 5$ AU in the 
heavier disk model. The scale height of this disk is given by 
$H/R \simeq 0.1$ (see paper 1). Figure~\ref{kinematic_viscosity1} shows 
that the kinematic viscosity at 5 AU is $\nu \simeq 3 \times 10^{15}$ cm$^2$ 
s$^{-1}$, which we adopt as the value for the vertical diffusion coefficient 
${\cal D}$. The diffusion time is given by
\begin{equation}
\tau_{\cal D} = \frac{H^2}{{\cal D}} = 625 \;\; {\rm yrs.}
\label{diff_time}
\end{equation}
We consider a simple chemistry (described in detail in 
section~\ref{model1}) involving a representative molecule and its ion, 
``m'' and ``m$^+$'', and a representative heavy metal and its ion, ``M'' 
and ``M$^+$''. In regions well shielded from X--rays, the recombination 
rate of free electrons is given by
\begin{equation}
\frac{d x[\mathrm{e}_{}^-]}{dt} = - k_1^{} x[\mathrm{e}_{}^-] x[{\rm m}_{}^+] N[{\rm H_2^{}}]
                       - k_2^{} x[\mathrm{e}_{}^-] x[{\rm M}_{}^+] N[{\rm H_2^{}}]
\label{recombination}
\end{equation}
where $k_1$ is the rate coefficient for recombination between molecular 
ions and electrons, $k_2$ is the equivalent for heavy metal ions, and 
$N[{\rm H_2^{}}]$ is the number density of hydrogen molecules. The two 
scenarios of interest are the metal--free case and the metal--dominated 
case.\\
\indent
In the metal--free case, where the electron fraction 
$x[\mathrm{e}_{}^-] = x[{\rm m_{}^+}]$, 
equation~(\ref{recombination}) becomes
\begin{equation}
\frac{d x[\mathrm{e}_{}^-]}{dt} = - k_1 x^2[\mathrm{e}_{}^-] N[{\rm H_2^{}}].
\label{recombination2}
\end{equation}
Assuming $N[{\rm H_2^{}}]$ remains constant, equation~(\ref{recombination2}) 
integrates to give the recombination time
\begin{equation}
\tau_{\tilde{\alpha}} = \frac{1}{k_1 N[{\rm H_2}]} \left[\frac{1}{x_{e^-}^f} -
\frac{1}{x_{e^-}^i}\right],
\label{tau_m+}
\end{equation}
where $x_{e^-}^i$ and $x_{e^-}^f$ are the initial and final values of 
$x_{e^-}$ during diffusion over a scale height. At $R=5$ AU, the electron 
abundance at the transition zone is $x_{e^-}^f = 5 \times 10^{-13}$, as shown 
in figure~7 of paper 1. The rate coefficient $k_1=\tilde{\alpha}$, 
$N[{\rm H}_2^{}] \simeq 5 \times 10^{12}$ cm$^{-3}$, and 
$T \simeq 100$ K (see figures~1 and 2 in paper 1). The recombination time is 
then $\tau_{\tilde{\alpha}} \simeq 15$ days. It is clear that diffusion is 
not expected to have a significant impact on the dead zone structure in this 
case.\\
\indent
The rate coefficient appropriate to the metal--dominated case is 
$k_2=\tilde{\gamma}$, which gives a recombination time of 
$\tau_{\tilde{\gamma}} \simeq$ 4200 yrs. Comparing this with the diffusion time 
$\tau_{\cal D} = 625$ yrs, it is clear that a protoplanetary disk with a 
significant population of gas--phase heavy metals should have the structure 
of the dead zone modified substantially by the action of turbulent diffusion.

\subsection{Oppenheimer \& Dalgarno model}
\label{model1}
\begin{figure}[t]
\includegraphics[width=.50\textwidth]{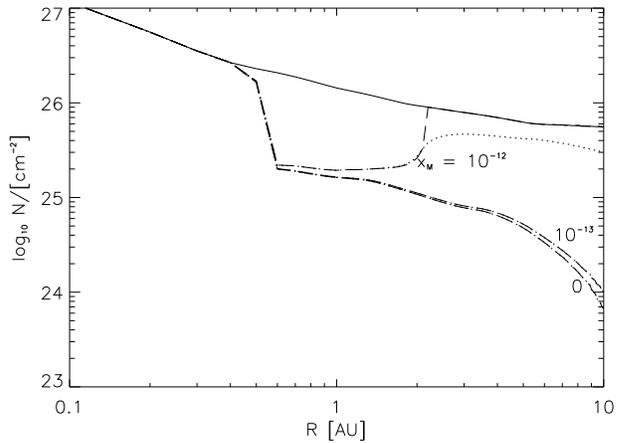}
\caption{- \texttt{model1(D)} - Column densities of the whole disk 
(solid line) and of the active zones (dashed and dotted lines) - 
refering to magnetic Reynolds numbers greater than 100 - for 
different values $x_{\mathrm{M}^{}}$. While the 
dashed lines refer to simulation with ${\cal D}  = \nu$, 
the dotted lines refer to simulations with ${\cal D} = 0$. 
The reference values for the rate coefficients are applied. The disk 
parameters are $\alpha = 10^{-2}$ and 
$\dot{M} = 10^{-7} \ \mathrm{M_{\odot} yr^{-1}}$.}
\label{column-model1D}
\end{figure}

\noindent
The underlying kinetic model of \texttt{model1D} involves two elements, 
five species/components, and four reactions. These species are free 
electrons ``$e^-_{}$", a representative molecule ``m", a heavy metal 
atom ``M", and their ionized counterparts ``m$^+_{}$" and ``M$^+_{}$". 
The rate coefficients for the ionisation of m and for dissociative 
\begin{table}[t]
\caption{Rate coefficients applied for the Oppenheimer \& 
Dalgarno model. The values are always used as reference values.}
\begin{center}
\begin{tabular}{lll}\hline\hline
$\zeta$ & $\zeta_{\mathrm{eff}}^{}$ & $\rm s^{-1}_{}$\\
$\tilde{\alpha}$ & $3 \times 10^{-6}_{}/ \sqrt{T}$ & $\rm cm^3_{} \ s^{-1}_{}$  \\
$\tilde{\beta}$  & $3 \times 10^{-9}_{}$ & $\rm cm^3_{} \ s^{-1}_{}$\\
$\tilde{\gamma}$ & $3 \times 10^{-11}_{}/ \sqrt{T}$ &$ \rm cm^3_{} \ s^{-1}_{}$  \\ 
\hline\hline          
\end{tabular}
\end{center}
\label{rates:oppenheimer}
\end{table}
recomination of m$^+_{}$ are given by $\zeta$ and $\tilde{\alpha}$, while 
the rate coefficients $\tilde{\beta}$ and $\tilde{\gamma}$ apply to the 
charge--transfer reaction between m$^+_{}$ and M and the radiative 
recombination of M$^+_{}$, respectively. The rate coefficients are listed 
in table~\ref{rates:oppenheimer} and were used in all \texttt{model1D} 
calculations unless stated otherwise. Results obtained in paper 1 showed 
that the dead zones obtained using \texttt{model1} were always smaller 
than those obtained for the complex models. We find it useful, however, 
to analyse \texttt{model1} and \texttt{model1D} as many of the features 
of this simple model can be helpful in understanding the more complex 
model.\\
\indent
The results obtained for \texttt{model1D} are presented in 
figure~\ref{column-model1D}, which shows the column density of the whole 
disk plotted as a function of radius using the solid line, and the column 
density of the active zone only using either dotted lines (for which 
${\cal D}=0$) or dashed lines (for which ${\cal D}=\nu$).\\
\indent
The first thing to note is that diffusive mixing makes essentially no 
difference to the size of the active zone when heavy metals are not 
included in the chemistry. This is illustrated by the lowest line plotted
in figure~\ref{column-model1D}, which appears as a dash-dotted line in the 
figure, but is actually a dashed line (representing the case with diffusion 
switched on) plotted over a dotted line (representing the case when diffusion 
is switched off). The reason for this result is simple: the recombination 
time when metals are absent is very much shorter than the turbulent mixing 
time. As a parcel of fluid moves from the active region of the disk toward 
the dead zone, free electrons will recombine with molecular ions rapidly, 
making no change to the ionisation fraction of the dead region.\\
\indent
Increasing the heavy metal elemental abundance from zero to 
$x_{\rm M} = 10^{-13}$ and switching diffusion on also makes no difference 
to the size of the dead zone. The addition of such a small fraction of metals 
does not prevent recombination between molecular ions and electrons being 
the dominant destruction process of the free electrons. This is illustrated by 
the second lowest line in figure~\ref{column-model1D}.\\
\indent
Increasing the heavy metal fraction to $x_{\rm M} = 10^{-12}$ results in a 
significant change in the structure of the dead zone when diffusion is 
switched on. Interior to $R \le 2$ AU there is almost no change in the size 
of the active zone because recombination between electrons and molecular ions 
is still the dominant process determining the ionisation fraction. Beyond 
$R \ge 2$ AU, however, the dead zone disappears completely when diffusion is 
switched on, whereas as a dead zone remains when $x_{\rm M} = 10_{}^{-12}$ and 
${\cal D}=0$. The reason for this difference is simple. The charge--transfer 
reaction between molecular ions and neutral heavy metals removes most of the 
molecular ions beyond $R \ge 2$ AU. This allows the recombination between metal 
ions and electrons to become the primary means by which free electrons are 
destroyed. The recombination time for this reaction, however, is much longer 
than for recombination between molecular ions and electrons, and is also longer 
than the turbulent mixing time. As a parcel of fluid moves from the active zone 
toward the dead zone, the time it takes is shorter than the time scale over which 
the free electrons are destroyed, thus increasing the free electron fraction in 
the dead zone and allowing it to become active. \\
\indent
In order to explore further the role played by the relative diffusion and 
recombination time scales, and to examine conditions under which the 
reaction--diffusion model deviates from the pure kinetic model, we have 
performed some numerical experiments in which we have artificially modified 
recombination times. We discuss these briefly below.\\[0.5em]
\noindent  
{\large {\bf Case: $x_{\rm M}^{} = 0$}}\\[.5em]
\noindent
In this case, \texttt{model1D} reduces to a three component system with two 
reactions. The molecular ion m$^+$ and the free electron $e^-$ are formed and 
destroyed by the same chemical reaction. Since m$^+_{}$ and $e^-_{}$ are mixed 
with the same diffusion velocity, it is simple to demonstrate how the mixing 
time scale, $\tau_{\cal D}$, and the molecular ion--electron recombination 
time scale, $\tau_{\tilde{\alpha}}$, affect the ionisation fraction $x[e^-_{}]$ 
(where $\tau_{\cal D} \propto 1/\nu$ and 
$\tau_{\tilde{\alpha}} \propto 1/\tilde{\alpha}$). We systematically changed 
the ratio $\tau_{\cal D}/\tau_{\tilde{\alpha}}$ by changing the rate coefficient 
$\tilde{\alpha}$ of the molecular ion -- electron recombination reaction. We 
find that the recombination time scale $\tau_{\tilde{\alpha}}$ has to be 
increased by three orders of magnitude before the results of the 
reaction--diffusion model \texttt{model1D} deviates from the pure kinetic 
model \texttt{model1}. This illustrates the basic point that reasonable 
diffusion rates will be unable to change the size of the dead zone obtained 
using the Oppenheimer \& Dalgarno model in the absence of heavy metals in 
the gas phase. \\[.5em]
\noindent  
{\large {\bf Case: $x_{\rm M}^{} \ne 0$}}\\[.5em]
\noindent
We performed a similar exercise to that just described, but included heavy 
metals with an elemental concentration of $x_{\mathrm{M}}^{} \ge 10_{}^{-12}$. 
This value causes the recombination of metal ions with free electrons to 
dominate over recombination with molecular ions beyond $R \ge 2$ AU. We used 
the reference values for the rate coefficients $\tilde{\alpha}$ and 
$\tilde{\beta}$ listed in table~\ref{rates:oppenheimer}, and modified 
systematically the recombination rate coefficient $\tilde{\gamma}$ between 
metals and electrons only. $\tau_{\tilde{\gamma}} \propto 1/\tilde{\gamma}$ 
now becomes the dominant recombination time scale. Figure~\ref{column-model1D} 
shows that the addition of metals with $x_{\mathrm{M}}^{} \ge 10^{-12}$ causes 
a substantial difference between the reaction--diffusion model and the pure 
kinetic model. We find that we need to decrease the value of 
$\tau_{\tilde{\gamma}}$ by two orders of magnitude in order that diffusion has 
no impact on the size of the dead zone.\\
\indent
One can consider this issue in reverse, and ask what changes need to be made 
to $\tau_{\tilde{\gamma}}$ in the pure kinetic model \texttt{model1} in order 
for it to give the same sized dead zone as \texttt{model1D} with 
${\cal D}=\nu$ and the standard value of $\tilde{\gamma}$. We find that if 
$\tau_{\tilde{\gamma}}$ is increased by two orders of magnitude in 
\texttt{model1} then the resulting dead zone is the same as obtained in the 
standard \texttt{model1D}. For this particular reaction network, this indicates 
that diffusion acts as an effective reduction in the dominant recombination 
rate coefficient.

\subsection{UMIST model}
\noindent
The underlying kinetic model in \texttt{model3D} was constructed by 
extracting all species and reactions from a given species set of 174 species 
and the UMIST database containing the elements H, He, C, O, N, S, Si, Mg, Fe. 
1965 reactions were extracted from the UMIST database. A full description is 
given in paper 1. Compared with the  previous model of Oppenheimer \& Dalgarno 
(\texttt{model1D}) there are now many pathways through which free electrons 
can recombine with molecular ions, each with their own associated time scale. 
Without knowing {\em a priori} which ion is dominant, it is difficult to make 
estimates of the relative mixing and recombination time scales. Indeed, our 
previous study presented in paper 1 suggests that there is more than one 
dominant ion that determines the ionisation fraction, complicating the picture 
further.\\
\indent
An additional complication is that the introduction of diffusion into the kinetic 
model can modify the chemical pathways, such that comparing the products of the 
pure kinetic models with those of the reaction--diffusion models becomes difficult. 
This effect manifests itself in our study by changing the identities of the molecular 
ions which are dominant in the transition region between dead and active zones.\\
\indent
We now discuss the results obtained using \texttt{model3D}, beginning with models 
for which the heavy metal, magnesium, was was neglected from the chemical network, 
before discussing the effects that including Mg has on the ionisation fraction and 
resulting dead zone structure. \\[.5em]
\noindent  
{\large {\bf Case: $x_{\rm Mg}^{} = 0$}}\\[.5em]
\noindent
We begin by discussing the results of \texttt{model3D} 
\begin{figure}[t]
\includegraphics[width=.50\textwidth]{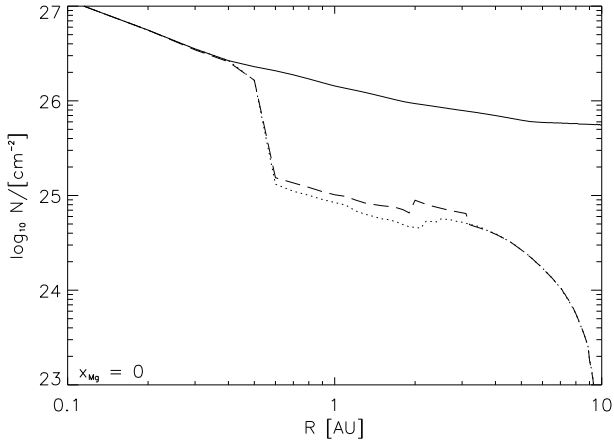}
\caption{- \texttt{model3(D)} - Column densities of the whole disk 
(solid line) and of the active zones (dashed and dotted lines) - 
refering to magnetic Reynolds numbers greater than 100 - for 
$x_{\mathrm{Mg}^{}} = 0$. While the dashed line refers to 
simulations with ${\cal D} = \nu$, the dotted line 
refers to simulation with ${\cal D} = 0$. The disk 
parameters are $\alpha = 10^{-2}$ and 
$\dot{M} = 10^{-7} \ \mathrm{M_{\odot} yr^{-1}}$.} 
\label{column-model3D-typeA}
\end{figure}
applied to the disk model defined by $\alpha = 10^{-2}$ and
$\dot{M} = 10^{-7} \ \mathrm{M_{\odot} yr^{-1}}$ before
discussing the results obtained in the disk model with
$\alpha = 5 \times 10^{-3}$ and
$\dot{M} = 10^{-8} \ \mathrm{M_{\odot} yr^{-1}}$.
As a general result we find that the location of the transition zone 
(which separates the active and the dead zones) is very similiar when 
comparing the results obtained for both disk models by applying a 
reaction--diffusion model with ${\cal D} = \nu $ and ${\cal D} = 0$, 
respectively. This is illustrated by figure~\ref{column-model3D-typeA} 
in which the column density of the whole disk is plotted using the 
solid line, the column density of the active zone obtained using the 
pure kinetic model is plotted using the dotted line, and the results 
of the reaction--diffusion model are plotted using the dashed line. We 
see that \texttt{model3D} with ${\cal D} = \nu $ produces a slightly 
larger active zone within $0.6 \le R \le 3 \ \mathrm{AU}$, but beyond 
this region the depth of the active zone is unaffected by diffusive 
mixing.\\
\indent
We have examined the abundances of the key molecular ions in the 
transition region between the active and dead zones. Typically the 
ionisation balance is not determined by a single dominant ion, but 
by a small number of ions. We find clear differences between the pure 
kinetic model and the reaction--diffusion model in those regions in 
radius where there is the greatest difference in the depth of the active 
zone between the two models. \\
\indent
HCNH$_{}^+$ and NH$_4^+$ are the most abundant ions in the neighbourhood 
of the transition zone between $1.8 \le R \le 3.0 \ \rm AU$ obtained for 
the pure kinetic model. It is important to note that the ratio between 
both ion concentrations does not change gradually in the vicinity of the 
transition zone, but instead there is a sharp transition: above the 
transition zone HCNH$_{}^+$ is more abundant than NH$_4^+$ by one order 
of magnitude. This ratio becomes inverted just below the transition zone 
by the same order of magnitude. Due to the significantly shorter 
recombination time scale, electrons recombine more efficiently with NH$_4^+$ 
than with HCNH$_{}^+$. Hence, NH$_4^+$ determines the location of the 
transition zone rather than HCNH$_{}^+$. \\
\indent
By contrast, the abundance of NH$_4^+$ is significantly lowered by applying 
the reaction--diffusion model with ${\cal D}=\nu$. Now, the more complex ion 
$\mathrm{H_4^{}C_2^{}N_{}^+}$ is two orders of magnitude the more abundant 
than the next most abundant ion $\mathrm{CH_3^{}OH_2^+}$. In addition, since 
mixing tends to reduce the concentration gradients of each species with height, 
$z$, the ratio between both ion concentrations 
$x[\mathrm{H_4^{}C_2^{}N_{}^+}]/x[\mathrm{CH_3^{}OH_2^+}]$ changes gradually 
when passing through the transition zone. In the same region between 
$1.8 \le R \le 3 \ \rm AU$ we found significant differences in the distribution 
of most of the species obtained for a reaction--diffusion model and the 
corresponding pure kinetic model. This is a clear indication that mixing may 
change the kinetics significantly.\\
\indent
We now consider the $\alpha = 5 \times 10_{}^{-3}$, 
$\dot{M} = 10_{}^{-8} \ \mathrm{M_{\odot} yr_{}^{-1}}$ disk model. The results 
for this model are shown in figure~\ref{column-model3D-typeR}. We first note 
that this disk model generates an intrinisically deeper dead zone in the outer 
regions of the disk beyond $R \ge 3$ AU, even though the ionisation rate due 
\begin{figure}[t]
\includegraphics[width=.50\textwidth]{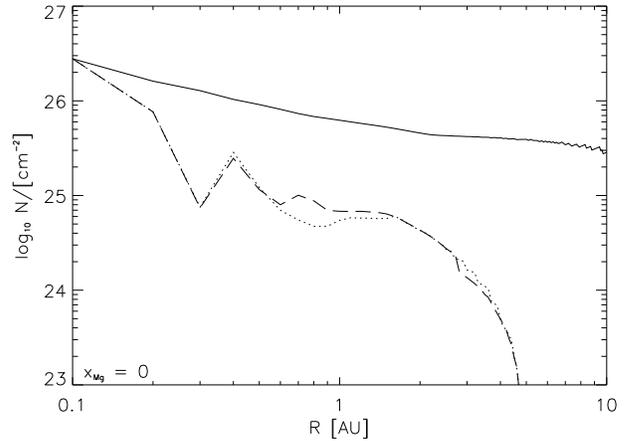}
\caption{- \texttt{model3(D)} - Column densities of the whole disk 
(solid line) and of the active zones (dashed and dotted lines) - 
refering to magnetic Reynolds numbers greater than 100 - for 
$x_{\mathrm{Mg}^{}} = 0$. While the dashed line refers to 
simulation with ${\cal D} = \nu$, the dotted line 
refers to simulation with ${\cal D} = 0$. The disk 
parameters are $\alpha = 5 \times 10^{-3}$ and 
$\dot{M} = 10^{-8} \ \mathrm{M_{\odot} yr^{-1}}$.} 
\label{column-model3D-typeR}
\end{figure}
to X--rays is higher due to the smaller surface density. This is a temperature 
effect that arises because the viscous heating is reduced because of the smaller 
$\alpha$ and ${\dot M}$ values. The recombination rates scale according to 
$k \propto 1/\sqrt{T}$, leading to a lower ionisation fraction.\\
\indent
We focus in particular on locations $0.6 \le R \le 1.1 \ \rm AU$ where the 
most significant differences in the vertical depth of the active zone appear 
when comparing the reaction--diffusion model and the corresponding pure kinetic 
model.\\
\indent
The most abundant ions produced by the pure kinetic model in the neighbourhood 
of the transition zone are $\mathrm{H_4^{}C_2^{}N_{}^+}$ and NH$_4^+$. These 
species are similarly abundant, but the recombination time for electrons with 
NH$_4^+$ is much shorter than with $\mathrm{H_4^{}C_2^{}N_{}^+}$, so NH$_4^+$ 
determines the position of the transition zone. The corresponding 
reaction--diffusion model again results in $x[\mathrm{NH_4^+}]$ dropping 
significantly so that it no longer controls the ionisation fraction in the 
neighbourhood of the transition zone. Instead, $\mathrm{H_4^{}C_2^{}N_{}^+}$ is 
the most abundant molecular ion, being one order of magnitude more abundant than 
the next most abundant ion $\mathrm{H_3^{}O_{}^+}$. This is again evidence that 
the presence of turbulent mixing in a protoplanetary disk can change the reaction 
kinetics. This leads to the intriguing possibility that such changes may in the 
future act as observational diagnostics for the presence of turbulence in disks.\\[.5em]
\noindent  
{\large {\bf Case: $x_{\rm Mg}^{} \ne 0$}}\\[.5em]
\noindent
We now consider the effect that including magnesium has on the models. We begin 
by discussing the results obtained for 
\begin{figure}[t]
\includegraphics[width=.50\textwidth]{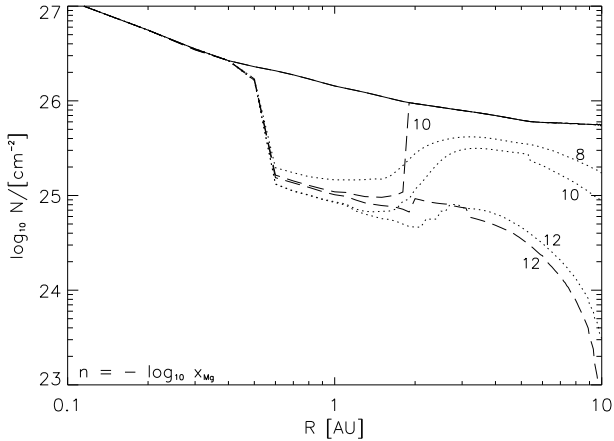}
\caption{- \texttt{model3(D)} - Column densities of the whole disk 
(solid line) and of the active zones (dashed and dotted lines) - 
refering to magnetic Reynolds numbers greater than 100 - for 
different values of $x_{\mathrm{Mg}^{}}$. While the dashed lines 
refer to simulations with ${\cal D} = \nu$, the dotted 
lines refer to simulations with ${\cal D} = 0$. 
Especially, for a reaction--diffusion model with 
${\cal D}_{\mathrm{eff}} = \nu$ no dead zones are observed above 
values $x_{\rm Mg}^{} \ge 10^{-8}$. The disk parameters are 
$\alpha = 10^{-2}$ and 
$\dot{M} = 10^{-7} \mathrm{M_{\odot} \ yr^{-1}}$.} 
\label{column-model3D-typeCDE}
\end{figure}
the disk model with $\alpha = 10_{}^{-2}$, 
$\dot{M} = 10_{}^{-7} \ \mathrm{M_{\odot} yr^{-1}}$. The results for 
\texttt{model3D} are shown in figure~\ref{column-model3D-typeCDE}. The 
dotted lines show the column density of the active zones obtained by 
the pure kinetic model for various values of $x_{\rm{Mg}}$. The dashed 
lines show the column density of the active zones obtained by the 
reaction--diffusion model. In general we find that mixing affects the 
size of the active zone significantly as soon as metals with 
concentrations above a threshold value are involved.\\
\indent
The addition of magnesium with elemental concentration 
$x_{\rm Mg}^{} = 10_{}^{-12}$ makes very little difference to the size 
of the dead zone, whether diffusion is switched on or not, as seen by 
comparing figures~\ref{column-model3D-typeA} and 
\ref{column-model3D-typeCDE}. This is because a low value of 
$x_{\rm Mg}^{}$ does not allow $\rm Mg_{}^+$ to become the dominant 
ion {\em via} charge--transfer reactions.\\
\indent
Increasing the magnesium elemental abundance to 
$x_{\rm Mg}^{} \ge  10_{}^{-10}$ changes the picture dramatically when 
diffusion is switched on. The pure kinetic models with these magnesium 
abundances included contain no regions in the disk beyond $R \ge 0.4$ AU 
where the disk is fully active, although the dead zone tends to shrink 
beyond $R \ge 2$ AU. In this region $\rm Mg_{}^+$ followed by ${\rm HCNH^+}$ 
are the most dominant ions, while in the region $0.4 \le R \le 2$ AU the 
${\rm NH_4^+}$ ion is dominant. The reaction--diffusion model \texttt{model3D} 
with ${\cal D}=\nu$ and $x_{\rm Mg}^{} = 10_{}^{-10}$ produces a dead zone of 
similar depth to the pure kinetic model between radii $0.4 \le R \le 2$ AU. 
The identity of the dominant molecular ion in this region changes, however, 
when diffusion is switched on. The abundances of ${\rm HCNH^+}$ and 
${\rm NH_4^+}$ are found to decrease significantly, and are replaced by the 
ion ${\rm H_4 C_2 N^+}$. The disk is rendered fully active beyond $R \ge 2$ AU 
where $\rm Mg_{}^+$ is the dominant ion and the recombination time increases 
above the mixing time.\\
\indent
\texttt{model3D} with ${\cal D}=\nu$ and $x_{\rm Mg}^{} = 10_{}^{-8}$ produces 
a fully active disk, in contrast to the corresponding pure kinetic model 
\texttt{model3}. For these higher magnesium abundances, ${\rm Mg}^{+}$ becomes 
the dominant ion through the charge--transfer reaction with molecular ions, and 
the abundance of ${\rm NH_4^+}$ is much reduced. The long electron recombination 
time associated with $\rm Mg_{}^+$ allows diffusive mixing to maintain a disk 
that is fully active.\\[.5em]
\noindent
We now consider the disk model with $\alpha = 5 \times 10_{}^{-3}$,
$\dot{M} = 10_{}^{-8} \ \rm M_{\odot}yr^{-1}$. The results are presented in 
figure~\ref{column-model3D-typeSTU}, where the column density of the active 
zones obtained using the pure kinetic models are plotted using dotted lines. 
Dashed lines represent the column density of active zones obtained using the 
reaction--diffusion model.\\
\indent
Adding magnesium with an elemental abundance $x_{\rm Mg} = 10_{}^{-12}$ results 
in very little change in the size of the dead zone whether diffusion is switched 
on or not (compare figures~\ref{column-model3D-typeR} and 
\ref{column-model3D-typeSTU}).\\
\indent
Increasing the magnesium elemental concentration to $x_{\rm Mg} = 10_{}^{-10}$ 
causes substantial changes to occur when diffusion is switched on. Interior to 
$R \le 0.7$ AU there is no change because the molecular ions ${\rm NH_4^+}$ and 
${\rm H_4C_2N^+}$ are dominant over ${\rm Mg}^+$ there, both with and without 
diffusive mixing. Between $0.7 \le R \le 3$ AU the disk becomes fully active 
when diffusion is switched on, but maintains a significant dead zone there in 
the pure kinetic model. The influence of ${\rm Mg}^+$ allows the recombination 
time to be increased so that mixing can render the disk active. Beyond $R \ge 5$ 
AU we obtain the surprising result that diffusive mixing causes the depth of the 
\begin{figure}[t]
\includegraphics[width=.50\textwidth]{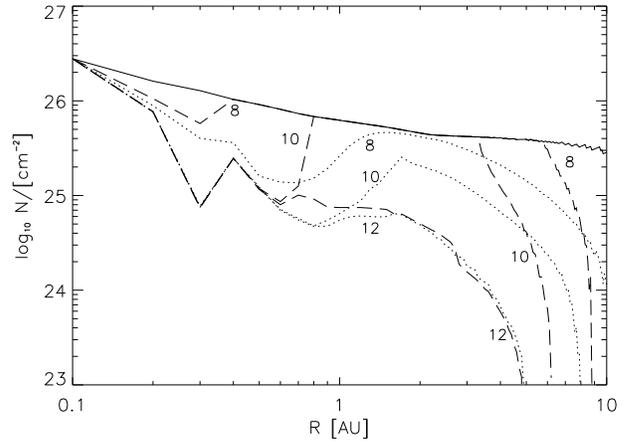}
\caption{- \texttt{model3(D)} - Column densities of the whole disk 
(solid line) and of the active zones (dashed and dotted lines) - 
refering to magnetic Reynolds numbers greater than 100 - for 
different values of $x_{\mathrm{Mg}^{}}$. While the dashed lines 
refer to simulations with ${\cal D} = \nu$, the 
dotted lines refer to simulations with ${\cal D}  = 0$. 
The disk parameters are $\alpha = 5 \times 10^{-3}$ and 
$\dot{M} = 10^{-8} \ \mathrm{M_{\odot} yr^{-1}}$.} 
\label{column-model3D-typeSTU}
\end{figure}
dead zone to increase rather than decrease. The reason for this is that the 
ionisation fraction at large radii in the pure kinetic model is always close to 
the critical value for the disk to be active. The addition of diffusion appears 
to have a diluting effect on the electron fraction in the active zone, which 
causes the dead zone to increase in size.\\
\indent
Increasing the magnesium elemental concentration to $x_{\rm Mg} = 10^{-8}$ causes 
further changes to occur when diffusion is switched on. The ${\rm Mg}^+$ ion is 
dominant throughout the disk, and the disk model is rendered entirely active 
between the radii $0.4 \le R \le 6$ AU. We again find that the dead zone becomes 
deeper at radii beyond $R \ge 8$ AU when diffusive mixing is effective. The reason 
is the same as given above: the ionisation fraction at these radii is always close 
to the critical value required to render the disk active. Diffusive mixing dilutes 
the electron fraction there, producing a deeper dead zone.

\section{Summary}
\label{summary}
We have presented the results of calculations that examine the ionisation fraction 
in protoplanetary disks models. Our models compute the chemical evolution of the gas, 
with X--rays from the central star being the primary source of ionisation, and also 
include vertical diffusion of the chemical species. This diffusion is designed to 
mimic the effects of turbulent mixing arising from the MHD turbulence that is thought 
to drive accretion in protostellar disks (e.g. Balbus \& Hawley 1991).\\
\noindent
The main findings of this work are:
\begin{itemize}
\item The simple Oppenheimer \& Dalgarno (1974) kinetic model produces smaller 
dead zones than a more complete kinetic model based on the UMIST database. This 
is true whether diffusionis switched on or not. 
\item All models which did not include heavy metals (magnesium) give rise to 
substantial dead zones. The inclusion of diffusion in these cases makes very 
little difference to the size and depth of the dead zone. This is because the 
recombination of molecular ions with electrons occurs too rapidly for turbulent 
mixing to be effective.
\item The addition of heavy metals (magnesium) to the reaction networks can 
give rise to dramatic changes in the sizes of dead zones when diffusion is 
included. This is because the recombination time between metal ions and 
electrons is orders of magnitudes longer than that between molecular ions and 
electrons, and is longer than the diffusive mixing time. The addition of metals, 
combined with charge--transfer reactions, removes most of the molecular ions and 
replaces them with the longer lived metal ions.
\item For the Oppenheimer \& Dalgarno (1974) network, the addition of heavy metals 
with fractional abundance $x_{\rm M}=10_{}^{-12}$ leads to the dead zone 
disappearing beyond $R \ge 2$ AU when diffusion is switched on. In the absence of 
diffusion a substantial dead zone exists in this region of this disk.
\item For the model based on the UMIST database, the addition of magnesium to the 
chemical network with a fractional abundance of $x_{\rm Mg}=10_{}^{-10}$ leads to 
the dead zone disappearing beyond $R \ge 2$ AU when diffusion is switched on. When 
the abundance of magnesium is increased to $x_{\rm Mg}=10^{-8}$ the dead zone 
disappears completely. Models in which diffusion is not present, but with the same 
magnesium abundances have substantial dead zones.
\item We find that the changes in the size of dead zones when diffusion is switched 
on do not only arise because free electrons are mixed into the dead zone, but also 
because the chemical pathways are modified by diffusion. One effect of this in our 
models is that the dominant ions near the transition zone between dead and active 
zones change their identity when diffusion is initiated. This raises the interesting 
possibility that there may be chemical signatures of turbulent diffusion that can 
act potentially as observational diagnostics of turbulence in protoplanetary disks.
\item For some disk models, the inclusion of diffusion can cause the dead zone 
depth to increase in the outer regions of the disk. This arises because these 
particlar models predict that the electron fraction is only slightly above the 
critical value for the disk to be active in these regions when diffusion is switched 
off. Switching diffusion on can dilute the electron fraction in these regions, 
leading to a deeper dead zone.
\end{itemize}
Overall, our models indicate that turbulence in protoplanetary disks can be a 
self--sustaining process, providing that certain criteria are met, and indicate 
that dead zones can be reduced or removed altogther through turbulent mixing 
processes. These criteria are:\\
({\it i}) 
There are sufficient magnesium atoms available in the gas phase so that recombination 
between magnesium ions and electrons becomes the dominant process by which the local 
ionisation fraction is determined. \\
({\it ii}) The turbulent diffusion time scale is shorter than the dominant recombination 
time on which free electrons are removed.\\
\indent
A potentially serious omission from our models is the existence of small dust grains. 
These are known to sweep up free electrons (and metal atoms) very efficiently, and 
when included with the interstellar abundance they substantially increase the size and 
depth of dead zones in protoplanetary disks (see Sano et al 2000; paper 1). It is very 
likely that these grains will need to be substantially depleted by grain growth before 
the effects of turbulent mixing described in this paper are realised, as the removal 
rate of electrons from the gas phase is too large to allow diffusion to be effective.\\
\indent
An interesting question is whether a disk that has a substantial dead zone, and which 
is subject to a temporary period of intense ionisation (by large X--ray flares of the 
type recently report by Feigelson et al. 2005, for example), can enter into a 
self--sustaining turbulent state in which the turbulent mixing maintains a fully active 
disk. This question is currently under investigation and will be the subject of a future 
publication. An equally interesting question is whether a disk with a substantial dead 
zone, sandwiched between active zones near the disk surface, can become globally active 
due to turbulent motions overshooting into the dead zone and transporting free electrons 
into it. This question cannot be addressed by calculations of the type presented in this 
paper, but their results suggest that this may be possible under certain circumstances.

\begin{acknowledgements}
\noindent 
This research was supported by the European Community's Research 
Training Networks Programme under contract HPRN-CT-2002-00308, 
"PLANETS". 
\end{acknowledgements}



\appendix

\section{}
\label{App1}
The diffusion flow ${\bf J}_i^a$ of the $i_{}^{\mathrm{th}}$ 
component may be expressed in various ways depending which 
reference velocity ${\bf v}_{}^a$ is used. The molar diffusion 
flow ${\bf J}_i^a$ is 
\begin{equation}
{\bf J}_i^a = n_i\left({\bf v}_i^{} - {\bf v}_{}^a \right).
\label{app:1}
\end{equation}
The reference velocities ${\bf v}_{}^a$ are defined as weighted 
averages of the velocity ${\bf v_i^{}}$ of the $i_{}^{\mathrm{th}}$ 
component
\begin{equation}
{\bf v}_{}^a = \sum_{i=1}^s a_i^{}{\bf v}_i, \ 
\left( \sum_{i=1}^s a_i^{} = 1 \right)
\label{app:2}
\end{equation}
with $a_i^{}$ as normalised weights.\\
\indent
Since the mass is conserved, the diffusion fluxes of the 
$s$ components are not independent. Specifying the reference 
velocity ${\bf v}_{}^a$ by the mean molar velocity 
${\bf v}_{}^m$
\begin{equation}
{\bf v}_{}^m = \sum_{i=1}^s x_i^{} {\bf v}_i^{}
\label{app:3}
\end{equation}
where $x_i^{}$ denotes the molar fraction of the 
$i_{}^{\mathrm{th}}$ component 
\begin{equation}
x_i^{} = \frac{n_i^{}}{n}, \ 
\left( \sum_{i=1}^s n_i^{} = n \right)
\label{app:4}
\end{equation}
the molar diffusion fluxes are constrained by
\begin{equation}
\sum_{i=1}^s{\bf J}_i^m = 0.
\label{app:5}
\end{equation}
\indent
There are several contributions to the diffusion flux. We assume 
that the most important contribution to the diffusion flux is 
caused by concentration gradients rather than due to pressure 
gradients, external forces, and coupled effects (cross--effects). 
Diffusion caused by concentration gradients is called 
``ordinary" diffusion. Considering ordinary diffusion processes 
only, we often use the phrase ``diffusion" instead of ``ordinary" 
diffusion. The (ordinary) diffusion is described by the 
phenomenological equation, which is well--known as Fick's first 
law for a binary system $s = 2$. By analogy to the molar diffusion 
flow for a binary system, we define an effective molar diffusion 
flow ${\bf J}_{\mathrm{eff},i}^m$ for the diffusion of the 
$i_{}^{\mathrm{th}}$ component in a mixture by 
\begin{equation}
{\bf J}_{\mathrm{eff},i}^m = 
- n \ {\cal D}_{\mathrm{eff},i}^{} \ \nabla x_i^{}.
\label{app:6}
\end{equation}
The effective binary diffusivity ${\cal D}_{\mathrm{eff},i}^{}$ 
for species $i$ in a mixture can be calculated from the diffusivity 
${\cal D}_{ij}^{}$ of the pair $i-j$ in a binary mixture by applying 
the so--called ``Stefan--Maxwell equation''. Especially, for systems 
in which all the ${\cal D}_{ij}^{}$ are the same 
${\cal D}_{\mathrm{eff},i}^{}$ is now given by
\begin{equation}
{\cal D}_{\mathrm{eff},i}^{} = {\cal D}_{ij}^{}.
\label{app:7}
\end{equation}
Because of the definition of the molar diffusion flow ${\bf J}_i^a$, 
see equation~(\ref{app:1}), the velocity of the $i_{}^{\mathrm{th}}$ 
component and the reference velocity ${\bf v}_{}^a$ are related to 
eachother. In particular, in a molar diluted system\footnote{i.e. 
for $n_i \ll 1, \ i=1,2,\dots, s - 1, \ n_s \sim 1$} we deal with, 
one can approximate the reference velocity by 
${\bf v}_i^a = {\bf v}_{}^m = 0$. Similiar applies for mass diluted 
systems with ${\bf v}_i^a = {\bf v}_n = 0$ where ${\bf v}_n$ denotes 
the $n_{}^{\mathrm{th}}$ component velocity. This is in agreement 
with the hydrostatic equilibrium assumption in $z$ direction. Finally, 
equation~(\ref{eq:3}) can be simplified by:
\begin{equation}
\frac{\partial n_i^{} }{\partial t}  = 
\frac{\ \partial}{\partial z} \left( n {\cal D}_{\mathrm{eff},i}^{} 
\frac{\ \partial}{\partial z} x_i^{} \right) + 
\sum_{j=1}^r \nu_{ij}^{} J_j^{}, \ (i=1,\dots, s)
\label{app:8}
\end{equation}
constraint by equation~(\ref{app:5}). Note that we simplified the 
symbol for the effective binary diffusivity for species $i$ in a mixture 
by using ${\cal D}$ instead of ${\cal D}_{\mathrm{eff},i}^{}$.\\
\indent
Because of the molar diluted system we consider and due to the fact 
that the most abundant components (which are molecular hydrogen and 
helium) do not contribute to the molar diffusion fluxes, the 
mass is conserved even by neglecting equation~(\ref{app:5}).

\end{document}